\documentstyle[preprint,aps]{revtex}

\begin{document}
\draft
\title{Compositeness condition at the next-to-leading order
            in the Nambu-Jona-Lasinio model }
\author{Keiichi Akama}
\address{Department of Physics, Saitama Medical College,
                     Kawakado, Moroyama, Saitama, 350-04, Japan}
\date{\today}
\maketitle

\begin{abstract}
We investigate the compositeness condition in four dimensions
	at the non-trivial order
	(next-to-leading order in $1/N$ with the number $N$ of the colors),
	taking the Nambu-Jona-Lasinio model as an example.
In spite of its non-renormalizability,
	the compositeness condition
	can be successfully derived and solved in compact forms.
The resultant problems are discussed.
\end{abstract}

Compositeness may be one of the most important concepts in physics
	to disclose the fundamental structure of the nature.
Needless to say, molecules, atoms, nuclei and hadrons are composites,
	and quarks, leptons, gauge bosons,
	and/or Higgs scalar also could be,
	as is supposed in some models \cite{comp} -- \cite{BHL}
	of the beyond-the-standard-model physics.
Furthermore composite states play essential roles in various phenomena,
	such as superconductivity, nuclear excitation etc..
Nevertheless many of the dynamical systems are complicated
	and hard to be solved except for some idealistic cases.
So far, extensive efforts have been devoted to this subject
	from both the particular and the general points of view.

Among them the ``compositeness condition" \cite{BHL} -- \cite{LT}
	in the field theory could be a powerful clue,
	and should be investigated more extensively.
This is the condition for a field in a model
	to be a composite of other fields,
	and expressed as
\begin{eqnarray}
	Z=0,	\label{Z=0}
\end{eqnarray}
where $Z$ is the renormalization constant of the field.
Under the condition (\ref{Z=0}), the composite field which is absent
	in the original bare-particle spectrum
	appears in the physical renormalized spectrum.
As is shown in \cite{cca},  
	a non-renormalizable four-fermion-interaction model is equivalent
	to a renormalizable Yukawa type model with the condition (\ref{Z=0}).
Then, what is urgent for the four-fermion model
	is to work out the compositeness condition.
In principle, it holds at any higher order.
In fact, Haymaker and Cooper \cite{HC} considered the Gross-Neveu model
\cite{GN}
	as a $Z=0$ limit of the two-dimensional $\sigma $ model,
	and investigated its higher-order effects.
In four dimensions, however, the compositeness condition
	has been derived and solved only at the lowest order \cite{cca},
	which merely reproduces the well known results
	in the earlier naive approach \cite{NJL}.
A true nontrivial developments would be obtained by going beyond it.
In recent years,
	people attempted to incorporate the higher order effects
	by using the ladder Schwinger-Dyson equation or
	the renormalization group equation
	\cite{MTY}, \cite{BHL}, \cite{LT}, \cite{higher}.
However, they are by nature incomplete because they take into account
	only particular classes of higher order diagrams.
The more rigorous and promising way is to use the equivalent Yukawa model
	with the compositeness condition mentioned above.
The purpose of this paper is to make a true nontrivial developments
	by deriving and solving the compositeness condition
	at the nontrivial order
	(next-to-leading order in $1/N$ with the number $N$ of the colors),
	and confirm the footing of the model.
Remarkably enough, the infinite sum of divergent diagrams
	can successfully be summed up to give compact forms,
	and its solution is very simple.

We consider the Nambu-Jona-Lasinio (NJL) model in its simplest form :
\begin{eqnarray}
	{\cal L}_{\rm N}=\overline \psi _0i\!\not\!\partial \psi _0
	+f_0|\overline \psi _{\rm 0L}\psi _{\rm 0R}|^2,
	\label{LNb}
\end{eqnarray}
where $\psi _0=(\psi _{01}, \psi _{02}, \cdots , \psi _{0N})$
	is a bare color $N$-plet fermion,
	$f_0$ is a bare coupling constant,
	the subscripts ``L" and ``R" indicate chiralities,
	and the subscript ``0" indicates that it is a bare quantity.
This model has $\rm U(1)_L\otimes U(1)_R$ chiral symmetry.
The system is equivalent to that described by the linearized
	Lagrangian \cite{KK}
\begin{eqnarray}
	{\cal L}'_{\rm N}=\overline \psi _0i\!\not\!\partial \psi _0
	+(\overline \psi _{\rm 0L}\phi _1\psi _{\rm 0R}+{\rm h.c.})
	-{1\over f_0}|\phi _1|^2,
	\label{L'Nb}
\end{eqnarray}
where $\phi _1$ is an auxiliary field.
In order to absorb the divergences we rescale the fields as
\begin{eqnarray}
	\psi _0= \sqrt {Z_\psi } \psi ,\ \ \ \ \
	\phi _1= \sqrt {{X}} \phi ,\ \ \ \ \
\end{eqnarray}
where $Z_\psi $ and $X$ are redundant constants, and
	$\psi $ and $\phi $ are the physical (renormalized) fields.
Then the Lagrangian (\ref{L'Nb}) becomes
\begin{eqnarray}
	{\cal L}'_{\rm N}=Z_\psi \overline \psi i\!\not\!\partial \psi
	+Z_\psi \sqrt {{X}}(\overline \psi _{\rm L}\phi \psi _{\rm R}+{\rm
h.c.})
	-{{{X}}\over f_0}|\phi |^2
	\label{L'NR}.
\end{eqnarray}

Now we compare this with the renormalizable Yukawa model
	for the elementary fermion $\psi _0$ and the elementary boson $\phi _0$
	with the following Lagrangian
\begin{eqnarray}
	{\cal L}_{\rm Y}=\overline \psi _0i\!\not\!\partial \psi _0
	+g_0(\overline \psi _{\rm 0L}\phi _0\psi _{\rm 0R}+{\rm h.c.})
	+|\partial _\mu \phi _0|^2
	-m_0^2|\phi _0|^2
	-\lambda _0|\phi _0|^4		\label{LY}
\end{eqnarray}
where $m_0$ is the bare mass of $\phi _0$,
	and $g_0$ and $\lambda _0$ are bare coupling constants.
To absorb the divergences of the quantum loop diagrams,
	we renormalize the fields, coupling constants and mass as
\begin{eqnarray}
&&	\psi _0= \sqrt {Z_\psi } \psi ,\ \ \ \ \
	\phi _0= \sqrt {Z_\phi } \phi ,\ \ \ \ \  \\
&&	Z_\psi \sqrt {Z_\phi }g_0= {Z_g} g,\ \ \ \ \
	Z_\phi ^2\lambda _0= Z_\lambda  \lambda ,\ \ \ \ \
	Z_\phi m_0^2= m^2+\delta m^2. \label{ZZdm}
\end{eqnarray}
where $\psi $, $\phi $, $g$, $\lambda $ and $m$ are the renormalized
	fields, coupling constants and mass, respectively, and
	$Z_\psi $, $Z_\phi $, $Z_g$, $Z_\lambda $ and $\delta m^2$ are
	the renormalization constants.
Then the Lagrangian ${\cal L}_{\rm Y}$ becomes
\begin{eqnarray}
&&	{\cal L}_{\rm Y}=Z_\psi \overline \psi i\!\not\!\partial \psi
	+Z_gg(\overline \psi _{\rm L}\phi \psi _{\rm R}+{\rm h.c.})
\cr &&\ \ \ \ \ \ \ \ \ \ \
	+Z_\phi |\partial _\mu \phi |^2
	-(m^2+\delta m^2)|\phi |^2
	-Z_\lambda \lambda |\phi |^4.		\label{LYR}
\end{eqnarray}

Now we can see that the Lagrangian (\ref{LYR}) of the Yukawa-type model
	coincides with the Lagrangian (\ref{L'NR}) of the NJL model, if
\begin{eqnarray}
&&	Z_\phi =Z_\lambda =0,	\label{CC}\\
&&	Z_\psi \sqrt {{X}}=Z_g  g,	\label{CY}\\
&&	{{X}}/f_0=-m^2-\delta m^2.	\label{Cf}
\end{eqnarray}
The conditions (\ref{CY}) and (\ref{Cf}) connect
	the parameters of the NJL and the Yukawa models,
	and lead to so-called ``the gap equation" of the NJL model.
On the other hand, (\ref{CC}) is the ``compositeness condition"
	which imposes relations among the parameters of the Yukawa model
	so that it reduces to the NJL model.
In terms of the bare parameters, (\ref{CC}) implies the limit
	$g_0\rightarrow \infty $ and $\lambda _0/g_0^4\rightarrow 0$.
Though some bare quantities diverge, renormalized ones are finite.
Thus the non-renormalizable NJL model is entirely equivalent
	to the renormalizable Yukawa model (with the cutoff fixed)
	under the compositeness condition (\ref{CC}) \cite{cca}.
Then what is urgent for the NJL model is to work out
	the compositeness condition.
For this purpose, we need to calculate $Z_\phi $ and $Z_\lambda $ in
	the Yukawa model and to solve $Z_\phi =Z_\lambda =0$ for $g$ and
$\lambda $.
In principle, we can perform it order by order to any higher order.

For an illustration, we start with the lowest order
	which reproduces (see below) the well known results
	in the earlier naive approach \cite{NJL}.
The divergences from the diagrams (a) and (b) in Fig.\ \ref{fig1}
	are absorbed by taking
\begin{eqnarray}
	Z_\phi =1-g^2NI,\ \ \ \ \
	Z_\lambda \lambda =\lambda -g^4NI,	\label{Z0}
\end{eqnarray}
where $I=\ln\Lambda ^2/16\pi ^2$ in the Pauli-Villars regularization,
	or $I=1/8\pi ^2(4-d)$ in the $d$-dimensional regularization.
The compositeness condition $Z_\phi =Z_\lambda =0$ with (\ref{Z0})
	is easily solved to give
\begin{eqnarray}
	g^2=\lambda =1/NI.	\label{sol0}
\end{eqnarray}
If $m^2<0$, then $\phi $ acquires a non-vanishing vacuum expectation value
$\langle\phi\rangle$
	and the chiral symmetry is broken, so that
	$\pi $=Im$\phi $ becomes the Nambu-Goldstone boson,
	and $\psi $ and $\sigma $=Re$\phi -\langle\phi\rangle$ respectively
acquire
the masses
\begin{eqnarray}
	m_\psi =\langle\phi\rangle/\sqrt {NI},\ \ \ \ \
	m_\sigma =2\langle\phi\rangle/\sqrt {NI}.	\label{m0}
\end{eqnarray}
The relations (\ref{sol0}) and (\ref{m0}) reproduce the well known results of
	the original NJL model \cite{NJL} with the chain approximation.
In this sense the method of the compositeness condition
	is trivial at this order.

Now we turn to the non-trivial order, which is the purpose of this paper.
First we note the fact that
	a $\psi $-loop insertion to a $\phi $-line or a $\phi $-vertex
	contributes a factor $O(g^2NI)\approx O(1)$ (because of (\ref{sol0})),
	and does not change the order of magnitude.
Then infinitely many higher order diagrams have the same order of magnitude
	and the naive expansion in $g$ and $\lambda $ fail.
A proper expansion is that in $1/N$ and $1/I$
	with the assignment $g^2=O(1/NI)$ and $\lambda =O(1/NI)$.
Here we retain the next-to-leading order in $1/N$
	and the leading order in $1/I$.
They consist of the infinitely many diagrams (c)--(g) in Fig.\ \ref{fig1},
	together with their counter terms
	for the sub-diagram divergences.

Now we derive the contribution to $Z_\phi $
	from the diagram (c) in Fig.\ \ref{fig1}.
Each $\psi $-loop inserted to the internal $\phi $-line with the momentum $p$
	contributes (in $d$-dimensional regularization)
\begin{eqnarray}
	g^2NIc_1(-p^2)^{1-\epsilon }	\label{Pi0}
\end{eqnarray}
with $\epsilon =(4-d)/2$ and $c_1\rightarrow 1$ as $d\rightarrow 4$.
If $l$ of $\psi $-loops (\ref{Pi0}) are inserted to the $\phi $-line,
	the self-energy part of $\psi $ with the momentum $k$
	is calculated to be
\begin{eqnarray}
	(g^2I)^{l+1}N^lc_2\!\not\!k/(-k^2)^{\epsilon (l+1)}+O(m^2)
\label{Sig}
\end{eqnarray}
with $c_2\rightarrow 1$ as $d\rightarrow 4$.
Then the diagram (c) in Fig.\ \ref{fig1} contribute to $Z_\phi $ the term
\begin{eqnarray}
	(-g^2I)^{l+2}N^{l+1}/(l+1)(l+2).
\end{eqnarray}
If we take into account the counter terms
	for the divergence of (\ref{Pi0}) and (\ref{Sig}),
\begin{eqnarray}
	(g^2I)^{l+2}N^{l+1}\sum _{k=0}^l\pmatrix{l\cr k}
	\left( {(-1)^k\over (k+1)(k+2)}-{(-1)^k\over (k+1)}\right)
	=-{(g^2I)^{l+2}N^{l+1}\over (l+1)(l+2)}.	\label{Pi1}
\end{eqnarray}
Collecting all together, we finally obtain the expression
\begin{eqnarray}
	Z_\phi =1-g^2NI-g^2I-{1\over N}(1-g^2NI)\log(1-g^2NI)+O({1\over
N^2}),\label{Z3}
\end{eqnarray}
where the logarithmic term appear
	through the infinite sum of (\ref{Pi1}) over $l$.
We can derive entirely the same expression as (\ref{Z3})
	also in the Pauli-Villars regularization.

Similarly the diagrams (d)--(g) in Fig.\ \ref{fig1}
	together with their counter terms for the sub-diagram divergences
	contribute to $Z_\lambda \lambda $ the terms
\begin{eqnarray}
&&	(d)\ \ -g^4I(g^2NI)^{l}/l(l+1),\ \ \ \
	(e)\ \ 10\lambda ^2I(g^2NI)^{l}/(l+1),\ \ \ \ \cr
&&	(f)\ \ -20\lambda g^2I(g^2NI)^{l}/(l+1),\ \ \ \
	(g)\ \ 10g^4I(g^2NI)^{l}/(l+1),
\end{eqnarray}
where $l$ is the number of the $\psi $-loops in the diagram.
Collecting all together, we obtain the expression
\begin{eqnarray}
&&	Z_\lambda \lambda =\lambda -g^4NI+8g^4I
	+{20(\lambda -g^2)^2I\over 1-g^2NI}
\cr &&	\ \ \
	-{1\over N}\left[ 2g^2(1-g^2NI)+20(\lambda -g^2)\right]
\log(1-g^2NI)+O({1\over N^3}).
\label{Z4}
\end{eqnarray}

The compositeness condition is obtained
	by putting $Z_\phi =Z_\lambda =0$ in (\ref{Z3}) and (\ref{Z4}).
Though it looks complicated at first sight,
	they are successfully solved giving the very simple solution
\begin{eqnarray}
	g^2={1\over NI}\left[ 1-{1\over N}+O({1\over N^2})\right] ,\ \ \ \ \ \ \
	\lambda ={1\over NI}\left[ 1-{10\over N}+O({1\over N^2})\right] .
\label{sol}
\end{eqnarray}
If the chiral symmetry is spontaneously broken
	the fermion $\psi $ and the scalar $\sigma $ acquire masses
\begin{eqnarray}
	m_\psi ={\langle\phi\rangle\over \sqrt {NI}}\left[ 1-{1\over
2N}+O({1\over
N^2})\right] ,\ \ \
	m_\sigma ={2\langle\phi\rangle\over \sqrt {NI}}\left[ 1-{5\over
N}+O({1\over
N^2})\right] .	\label{m}
\end{eqnarray}
This means that the famous ratio $m_\sigma /m_\psi =2$
	in the lowest-order NJL model should be corrected as
\begin{eqnarray}
	{m_\sigma \over m_\psi }=2\left[ 1-{9\over 2N}+O({1\over N^2})\right] .
\label{r}
\end{eqnarray}

The results are somewhat unpleasant for small $N$,
	in particular, for $N$=3 of our practical interest.
The next to leading contributions to $Z_\lambda $,
	and accordingly, those to $\lambda $ are too large.
Furthermore $\lambda $ is negative, and hence the Higgs potential becomes
unstable.
Its true stability may depend on the effective potential at large $\phi $.
The difference, however, comes from the convergent diagrams,
	and is suppressed by a factor of $1/I$
	compared with the Higgs potential itself.
The non-leading effects in $1/I$ may improve the situation.
A possible way to go around the difficulty
	may be to take into account the gauge interactions
	and/or higher-dimensional interactions \cite{hid}.
On the other hand, it may be too simple-minded
	to take the cutoffs of $\psi $ and $\phi $ as equal.
Introduction of the two cutoff scales could improve the situation \cite{2c}.
These possible improvements are now under consideration.

We have derived and solved the compositeness condition
	at the non-trivial order,
	which has turned out to have large contributions at $N=3$.
It is straightforward to extend the investigation to
	more realistic cases of, for example,
	${\rm SU(2)_L\otimes SU(2)_R}$ for hadron physics,
	or ${\rm SU(2)_L\otimes U(1)_Y}$ for electroweak theory.
In the realistic model, we should take into account
	the effects of gauge bosons also.
The present method is also applicable
	to the models of composite fermions
	and composite gauge bosons \cite{compG},
	and to the induced gravity theory (pregeometry) \cite{preg}.
Another challenge may be to apply it to the QCD bound states.
Recently some people used the compositeness condition
	as a boundary condition
	of the renormalization group equation \cite{BHL}.
It may also require the higher order consideration as is given here.
More thorough investigations including further higher orders
	would lead to deeper insights into the compositeness.

The author would like to thank Professors
H.~Terazawa,
A.~J.~Macfarlane,
R.~W.~Haymaker,
I.~Ichinose,
M.~Suzuki,
T.~Hattori,
T.~Matsuki,
H.~Murayama,
H.~Nakano,
M.~Yasu\`e,
and
Y.~Watabiki
for stimulating and critical discussions.


\newpage

\begin{figure}
\caption{
The diagrams which are leading ((a), (b))
	and next-to-leading ((c)--(g)) in $1/N$
	and leading in $1/I$.
The solid and dashed lines indicate $\psi $ and $\phi $ propagators,
respectively.
The big-dotted line indicates the $\phi $ propagator with
	a number of one-$\psi $-loop insertions.
The blob indicates a planar or a crossed one-$\psi $-loop insertion
	at the vertex.
}
\label{fig1}
\end{figure}

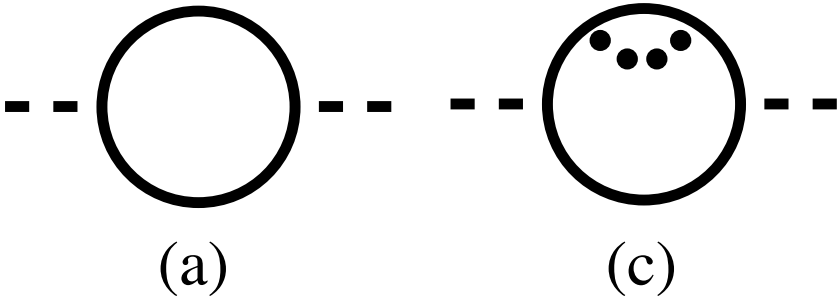
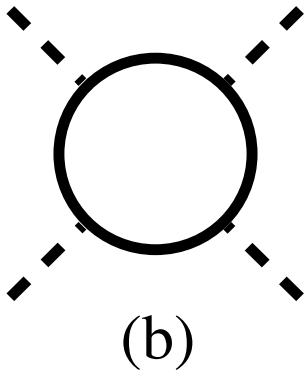
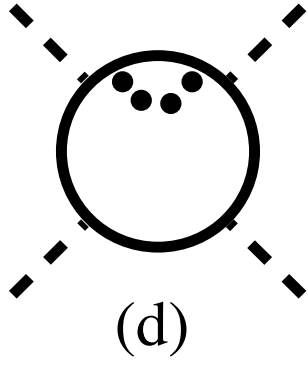
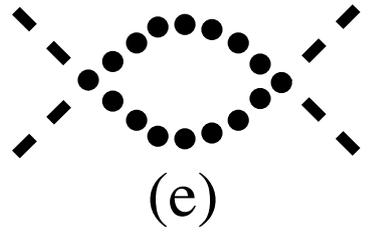
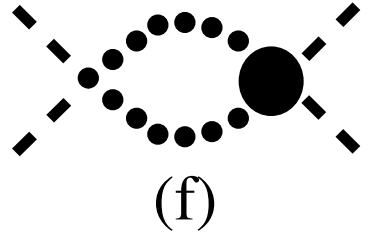
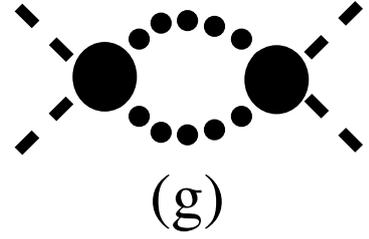
\end{document}